\documentclass[prd,twocolumn,nofootinbib,superscriptaddress]{revtex4-2}
\pdfoutput=1
\input epsf
\usepackage{graphics}
\usepackage{amsmath,amssymb}
\usepackage{color}
\usepackage{dcolumn}
\usepackage{hyphenat}
\usepackage{multirow}

\usepackage{graphicx}

\def\be{\begin{equation}}
\def\ee{\end{equation}}
\def\ba{\begin{eqnarray}}
\def\ea{\end{eqnarray}}

\newcommand{\beqa}{\begin{eqnarray}}
\newcommand{\eeqa}{\end{eqnarray}}

\newcommand{\beq}{\begin{equation}}
\newcommand{\eeq}{\end{equation}}

\newlength{\tskip}
\newlength{\colwidth}

\begin{document}

\title{Hints of Primordial Magnetic Fields at Recombination and Implications for the Hubble Tension}

\author{Karsten Jedamzik} 
\email[]{karsten.jedamzik@umontpellier.fr}
\affiliation{Laboratoire de Univers et Particules de Montpellier, UMR5299-CNRS, Universite de Montpellier, 34095 Montpellier, France}

\author{Levon Pogosian} 
\email[]{levon\_pogosian@sfu.ca}
\affiliation{Department of Physics, Simon Fraser University, Burnaby, BC, V5A 1S6, Canada}

\author{Tom Abel}
\email[]{tabel@stanford.edu}
\affiliation{Kavli Institute for Particle Astrophysics and Cosmology, Stanford University, 452 Lomita Mall, Stanford, CA 94305, USA}
\affiliation{Department of Physics, Stanford University, 382 Via Pueblo Mall, Stanford, CA 94305, USA}
\affiliation{SLAC National Accelerator Laboratory, 2575 Sand Hill Road, Menlo Park, CA  94025, USA}

\begin{abstract}
Primordial Magnetic Fields (PMFs), long studied as relics of the early Universe, accelerate recombination and have been proposed as a way to relieve the Hubble tension.  However, previous studies relied on simplified toy models. Here we use recent evaluations of recombination with PMFs, incorporating full magnetohydrodynamic (MHD) simulations and detailed Lyman-alpha radiative transfer, to test PMF-enhanced recombination ($b\Lambda$CDM) against observational data from the cosmic microwave background (CMB), baryon acoustic oscillations (BAO), and Type Ia supernovae (SN). Focusing on non-helical PMFs with a Batchelor spectrum, we find a preference for present-day total field strengths of approximately 5-10 pico-Gauss. Depending on the dataset combination, this preference ranges from mild ($\sim 1.8\sigma$ with Planck + DESI) to moderate ($\sim 3\sigma$ with Planck + DESI + SH0ES-calibrated SN) significance. The $b\Lambda$CDM has Planck + DESI $\chi^2$ values equal to or better than $\Lambda$CDM while predicting a higher Hubble constant. Future high-resolution CMB temperature and polarization measurements will be crucial for confirming or further constraining PMFs at recombination. Field strengths of 5-10 pico-Gauss align closely with those required for cluster magnetic fields to originate entirely from primordial sources, without the need for additional dynamo amplification.
\end{abstract}

\maketitle

\section{Introduction}

With cosmology entering the era where multiple independent datasets are able to constrain key aspects of the cosmological model with precision comparable to that of the cosmic microwave background (CMB), tensions between certain datasets have emerged in the context of the $\Lambda$ Cold Dark Matter ($\Lambda$CDM) model. Most prominent among them is ``the Hubble tension'' -- a $5\sigma$ discrepancy between the value of the Hubble constant, $H_0 = 67.36 \pm 0.54$ km/s/Mpc, inferred from the Planck CMB data~\cite{Planck:2018vyg}, and $H_0 = 73.04 \pm 1.04$ km/s/Mpc measured by the SH0ES collaboration using supernovae Type Ia (SN) calibrated on Cepheid stars observed by the Hubble Space Telescope (HST)~\cite{Riess:2021jrx,Riess:2022mme}. SN calibrations using Tip of the Red Giant Branch (TRGB) stars, J-Region Asymptotic Giant Branch (JAGB) stars, and Cepheids from the James Webb Space Telescope (JWST) also yield larger $H_0$~\cite{Freedman:2024eph}, albeit with larger uncertainties, with the uncertainties expected to shrink considerably as more JWST data becomes available. A lesser tension concerns measurements of the matter clustering amplitude quantified by the parameter $S_8$. The Planck-CMB-inferred value is $S_8 = 0.832 \pm 0.013$, while the joint analysis of galaxy counts and galaxy weak lensing data~\cite{Kilo-DegreeSurvey:2023gfr} from the Dark Energy Survey Year 3 (DES-Y3)~\cite{DES:Y3} and Kilo-Degree Survey (KiDS-1000)~\cite{KiDS:1000} yields $S_8 = 0.790^{+0.018}_{-0.014}$.  Another minor tension emerged recently between the value of the matter density fraction, $\Omega_m$, obtained from uncalibrated SN luminosities and that from the Baryon Acoustic Oscillations (BAO) measurements. The Pantheon+ (PP) SN data yields $\Omega_m = 0.334 \pm 0.018$~\cite{Brout:2022vxf}, in good agreement with DES-Y5~\cite{DES:2024jxu} and Union3~\cite{Rubin:2023ovl} SN datasets, while the BAO measurements by the Dark Energy Spectroscopic Instrument (DESI) give $\Omega_m = 0.295 \pm 0.015$~\cite{DESI:2024mwx}. 

The Hubble tension generated significant interest in extensions of the $\Lambda$CDM model~\cite{Abdalla:2022yfr}. Of special interest was the realization that Primordial Magnetic Fields (PMFs), if present in the pre-recombination plasma, would generate baryon inhomogeneities and speed up the recombination process~\cite{Jedamzik:2013gua}, bringing the CMB-deduced value of $H_0$ closer to that found from distance-ladders~\cite{Jedamzik:2020krr}. These findings, reproduced by other groups~\cite{Thiele:2021okz,Rashkovetskyi:2021rwg}, were based on a simple toy model of baryon clumping. It was also found~\cite{Lee:2022gzh,Lynch:2024gmp,Lynch:2024hzh,Mirpoorian:2024fka} that recombination histories similar to those predicted by this toy model were favoured by the combination of CMB and DESI BAO data, while yielding a larger $H_0$. Unlike many specially designed solutions to the Hubble tension, PMFs have been a subject of continuous study over many decades as the possible source of observed galactic and cluster magnetic fields~\cite{Widrow:2002ud}. Interest in PMFs increased with tentative evidence for magnetic fields in the extragalactic space from non-observation of GeV $\gamma$-ray halos around TeV blazars~\cite{Neronov:2010gir} and the synchrotron emission from a few-Mpc-long ridge connecting two merging clusters of galaxies~\cite{Govoni_2019}. However, with non-primordial astrophysical explanations of these low-redshift observations being difficult to rule out~\cite{Broderick:2018nqf}, the only sure way to prove the primordial origin of cosmic magnetic fields would be to find their signatures in the CMB. For a long time, CMB-based studies of PMFs have only led to $\sim$ nano-gauss (nG) upper bounds on the comoving field strength~\cite{Planck:2015zrl,Zucca:2016iur}, until the realization that the effects of baryon clumping on recombination may strengthen such upper limits by up to two orders of magnitude~\cite{Jedamzik:2018itu}.

Another significant development happened in the past four years that, together with baryon clumping, has taken PMF studies to a new level. Finding viable models of primordial magnetogenesis proved challenging, with inflationary magnetogenesis requiring postulating new physics to break the conformal invariance of electromagnetism~\cite{Turner:1987bw,Ratra:1991bn,Durrer:2013pga}, while PMFs generated in phase transitions~\cite{Vachaspati:1991nm,Vachaspati:2020blt}, until recently, were not expected to leave observable signatures in the CMB~\cite{Wagstaff:2014fla}. The prediction for phase-transition-generated PMFs depends crucially on the evolution of the field between the epoch of magnetogenesis and recombination. Leaning on analogies with hydrodynamics, Ref.~\cite{Banerjee:2004df} formulated such an evolutionary model. However, recent detailed studies~\cite{Hosking:2020wom} of freely decaying MHD turbulence show that the analogy with hydrodynamics is not perfect. The authors of~\cite{Hosking:2020wom} proposed (see also~\cite{Brandenburg:2014mwa,Zrake:2014mta}) that even non-helical magnetic fields undergo an inverse cascade governed by the conservation of a new invariant named ``Hosking integral'', and that the time-scale of evolution is not governed by the Alfven time-scale but the much slower magnetic reconnection time-scale. The first claim was then subsequently confirmed by \cite{Zhou:2022xhk}, whereas the second claim has not been found to be correct, although a significantly slower evolution time-scale was found in~\cite{Brandenburg:2024tyi} than that previously assumed. It is important to note that such studies cannot currently be performed for realistic Lundquist numbers relevant to the early universe plasma, but they indicate that phase-transition-generated PMFs may well evolve to produce enough PMF strength to relieve the Hubble tension and explain extragalactic magnetic fields~\cite{Hosking:2022umv}.

Earlier studies of the PMF effects on recombination and their impact on CMB anisotropies were based on simple toy models, which moreover did not have a straightforward relation to the final magnetic field strength. With future CMB temperature and polarization experiments promising to distinguish deviations from standard recombination at a few-percent level~\cite{Lee:2022gzh,Lynch:2024gmp,Lynch:2024hzh,Mirpoorian:2024fka}, a need for more detailed theoretical modelling of the recombination process in the presence of PMFs emerged. Significant progress in this direction was made over the past two years. Extensive 3D MHD simulations combined with Lyman-$\alpha$ photon transport were performed in~\cite{Jedamzik:2023rfd} to determine the redshift evolution of the ionized fraction $x_e(z)$ from $z=4500$ to $z=10$ for non-helical PMFs with a Batchelor spectrum. In this paper, for the first time, we use the MHD-derived $x_e(z)$ in the appropriately modified standard Boltzmann code {\tt CAMB}~\cite{Lewis:1999bs} to compute CMB anisotropy spectra and derive constraints on the PMF and the Hubble constant from the available data.

\section{Results}

\begin{table*}[!htbp]
\centering
\begin{tabular}{c|c|c|c|c|c}
 & $\chi^2_{b\Lambda{\rm CDM}}- \chi^2_{\Lambda{\rm CDM}}$  & \multicolumn{3}{c|}{$b_{\rm pmf}$ [nG]} \\
 & & median \& 68\% CI & 95\% CI & 99.7\% CI \\
\hline \hline
PL & $-1.7$ &  $0.0023^{+0.0048}_{-0.0022}$ & $<0.0075$
 & $<0.015$ \\
PL+DESI & $-4.7$ &  $0.0042^{+0.0021}_{-0.0028}$ & $<0.0083$ & $<0.013$ \\
PL+DESI+PP &  $-3.83$ & $0.0038^{+0.0017}_{-0.0028}$ & $<0.0079$ & $< 0.011$\\
PL+DESI+ACT & $-5.3$  & $0.0030^{+0.0012}_{-0.0021}$ & $<0.0067$ & $<0.008$\\
PL+DESI+SPT & $-5.2$ & $0.0036^{+0.0015}_{-0.0023}$ & $<0.0075$ & $<0.010$ \\
PL+DESI+PP+$M_b$ & $-15.25$ &  $0.0096^{+0.0029}_{-0.0036}$ & $[0.0049, 0.016]$ & $[0.0024, 0.024]$\\
PL+DESI+ACT+PP+$M_b$ & $-11.8$  & $0.0064\pm 0.0021$ & $[0.0025, 0.012]$ & $<0.016$\\
PL+DESI+SPT+PP+$M_b$ & $-17.8$ & $0.0074^{+0.0018}_{-0.0027}$ & $[0.0031, 0.013]$ & $[0.0019, 0.016]$ \\
 \hline \\
\end{tabular}
\caption{\label{tab:deltachi2} 
Differences in $\chi^2$ between $b\Lambda$CDM and $\Lambda$CDM fits to Planck (PL), and combinations of PL with DESI BAO, uncalibrated Pantheon+ (PP) SN, and PP calibrated using the SH0ES value of the absolute magnitude (PP+$M_b$), along with the median values and the 68\%, 95\% and 99.7\% credible intervals (CI) for the $z=10$ PMF strength $b_{\rm pmf}$. While the lower credibility bound is never strictly zero for a non-negative parameter, we set it to zero when it is below $0.0002$ nG, {\it i.e.} below the accuracy threshold of our simulations. The corresponding PMF strengths at the epoch of recombination are about an order of magnitude larger (See Table~\ref{tab:brec})).}
\end{table*}

\begin{table*}[!tbp]
\centering
\begin{tabular}{c|c|c|c}
 & $\Lambda$CDM PL+DESI & $b\Lambda$CDM PL+DESI & $b\Lambda$CDM PL+DESI+PP+$M_b$  \\
\hline \hline
$b_{\rm pmf}$ [nG] & - & $0.0042^{+0.0021}_{-0.0028}$ & $0.0096^{+0.0029}_{-0.0036}$\\
$H_0$ [km/s/Mpc] & $67.88 \pm 0.37$ & $68.52^{+0.54}_{-0.62}$ & $69.93^{+0.53}_{-0.66}$\\
$\Omega_m$ & $0.3066 \pm 0.0049$ & $0.3024 \pm 0.0055$ & $0.2917 \pm 0.0048$\\
$S_8$ & $0.8154 \pm 0.0090$ & $0.8154 \pm 0.0090$ & $0.8095 \pm 0.0087$\\
$\chi^2_{\rm Planck}$ & $10973.9$ & $10971.6$ & $10977.6$\\
$\chi^2_{\rm DESI} $ & $16.55$    & $14.0728$& $12.91$\\
$\chi^2_{\rm Planck+DESI}$ & $10990.4$ & $10985.7$ & $10990.5$\\
 \hline \\
\end{tabular}
\caption{\label{tab:params} 
The median values and the 68\% credible intervals of $b_{\rm pmf}$, $H_0$, $S_8$ and $\Omega_m$ obtained from PL+DESI and PL+DESI+PP+$M_b$, along with the best-fit $\chi^2$ values for PL, DESI and PL+DESI.}
\end{table*}

Our study considers non-helical fields with a Batchelor spectrum. We present our constraints on PMFs in terms of the total ({\it i.e.} integrated over all scales) root-mean-square comoving field strength $b_{\rm pmf}$ at $z=10$ corresponding to the final time of the MHD simulations~\cite{Jedamzik:2023rfd}. The corresponding field strength at recombination, evaluated at the peak of the visibility function ($z \approx 1090$), is approximately a factor of ten larger. This is due to additional damping that takes place shortly after recombination as the plasma transforms from highly viscous to turbulent state. Weaker, logarithmic damping occurs between $z \approx 100$ and $z = 0$~\cite{Banerjee:2004df}, which means the present-day PMF is comparable to $b_{\rm pmf}$, but about 17\% smaller. We refer to $\Lambda$CDM with a PMF as the $b\Lambda$CDM model.

Details of the computations of the ionization history and the final field strength, such as the resolution, the simulated modes and initial normalization, are given in Sec.~\ref{sec:MHD} and Sec.~\ref{sec:ionization} of Methods. In deriving the constraints on PMFs, we used the mean ionization histories obtained by averaging over 5 MHD realizations at each magnetic strength. This method does not account for the considerable sample variance around the mean predictions. To estimate just how much of a difference this makes, we separately performed a parameter fit using simulations with randomly generated ionization histories based on the MHD-derived covariance. The details of this investigation are presented in Sec.~\ref{sec:ensemble} of Methods, where we show that, as expected, accounting for the sample variance increases the uncertainty in $b_{\rm pmf}$, while the uncertainties in $H_0$ and other cosmological parameters remain largely unchanged. 

Table~\ref{tab:deltachi2} summarizes the posterior median values and credible intervals of $b_{\rm pmf}$ obtained from all the different data combinations considered in this work, along with the improvements in the combined $\chi^2$ relative to the corresponding $\Lambda$CDM model fits. The datasets and the model parameter priors are described in Sec.~\ref{sec:datasets} of Methods. We find that the Planck (PL) data by itself shows $\sim 1\sigma$ preference for $b_{\rm pmf}$, yielding $0.0023^{+0.0048}_{-0.0022}$ nG and only a modest improvement of $-1.7$ in the $\chi^2$ after adding one new model parameter. 

Combining Planck with DESI BAO shows a minor preference for a non-zero PMF, $b_{\rm pmf} = 0.0042^{+0.0021}_{-0.0028}$ nG, while reducing the combined PL+DESI $\chi^2$ by $4.7$. This is driven by the well-known preference of the DESI Year 1 data for a larger product of the Hubble constant and the sound horizon at baryon decoupling $r_{\rm drag} h$~\cite{DESI:2024mwx} compared to the Planck $\Lambda$CDM derived value (see also Table~\ref{tab:params_lcdm}). Accounting for the PMFs allows for larger CMB-derived $r_{\rm drag} h$ values, bringing CMB and BAO in better agreement with each other~\cite{Pogosian:2024ykm}.

Combining Planck and DESI data with uncalibrated SN from Pantheon+ (PP) reduces the preference for PMFs compared to the case without PP, due to PP favouring a larger $\Omega_m$, while still improving the $\chi^2$ by $3.8$ relative to $\Lambda$CDM. However, further combining it with the distance-ladder-determined SN magnitude $M_b$ from SH0ES (``the SH0ES prior''), yields $b_{\rm pmf} =0.0096^{+0.0029}_{-0.0036}$ nG with $H_0=69.93^{+0.53}_{-0.66}$ km/s/Mpc, while reducing the combined $\chi^2$ by $15.25$. As one can see from Table~\ref{tab:deltachi2}, $b_{\rm pmf} =0$ is excluded beyond $3\sigma$ in this case.

\begin{figure}[htbp!]
\includegraphics[width=0.48\textwidth]{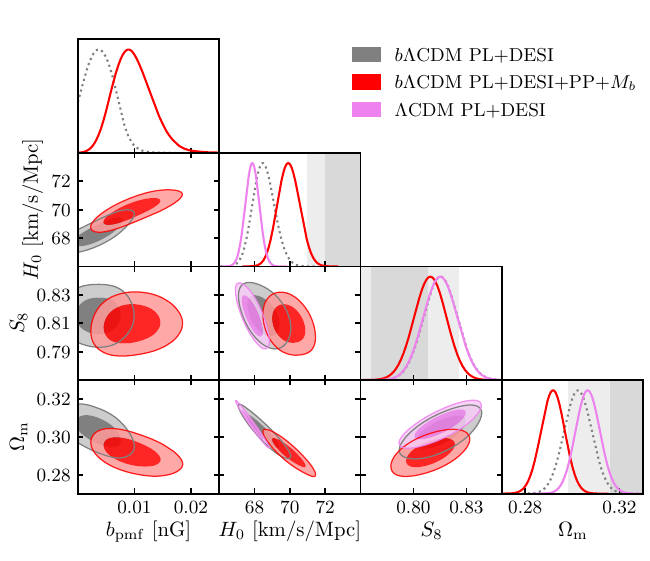}
\caption{\label{fig:pmf4} 
The marginalized $68$\% and $95$\% CL for mean $x_e(z)$ runs for Planck+DESI, Planck+DESI+PP+Mb, and the $\Lambda$CDM fit to Planck+DESI. Shown with grey bands are the 68\% and 95\& CL bands for the $H_0$ measurement by SH0ES~\cite{Riess:2021jrx}, $S_8$ from DES+KiDS~\cite{KiDS:1000}, and $\Omega_m$ from PP~\cite{Brout:2022vxf}.}
\end{figure}

\begin{figure}[htbp!]
\includegraphics[width=0.48\textwidth]{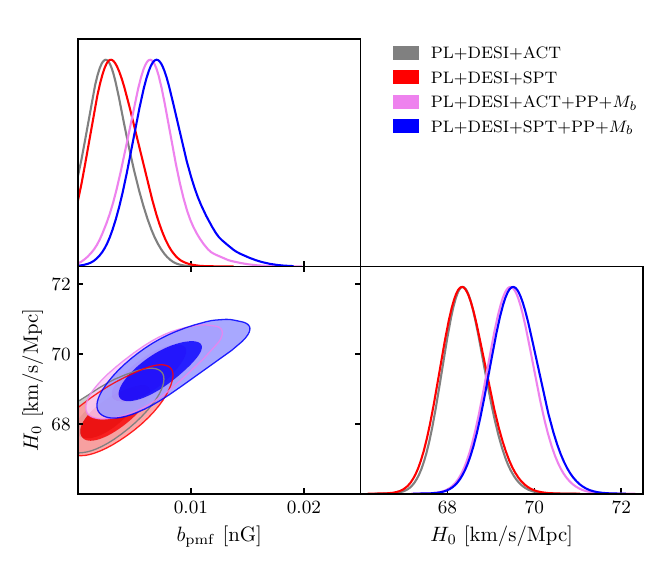}
\caption{\label{fig:actspt} 
The marginalized $68$\% and $95$\% CL for $b_{\rm pmf}$ and $H_0$ from fits to Planck+DESI+ACT and Planck+DESI+SPT with and without the ``SH0ES prior'', PP+$M_b$.}
\end{figure}

\begin{figure}[htbp!]
\includegraphics[width=0.48\textwidth]{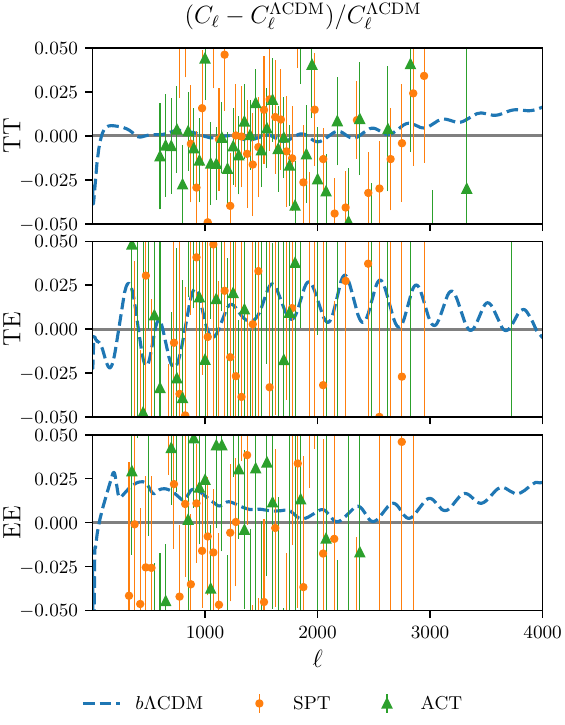}
\caption{\label{fig:residuals} 
The CMB TT, TE and EE residuals for the best-fit Planck+DESI+PP+$M_b$ $b\Lambda$CDM model relative to the $\Lambda$CDM best-fit to Planck+DESI, along with the ACT and SPT band-powers showing the mean values and the one-standard-deviation error bars.}
\end{figure}

The reduction in the PL+DESI+PP+$M_b$ $b\Lambda$CDM $\chi^2$ is dominated by the improvement in the SH0ES fit, bringing the disagreement with the SH0ES-determined value of $H_0=73.04 \pm 1.04$ km/s/Mpc down to $2.7\sigma$. We note, however, that the corresponding $b\Lambda$CDM value of $M_b=-19.364^{+0.016}_{-0.018}$ is $3.5\sigma$ different from the SH0ES determination of $M_b=-19.253 \pm 0.027$. {\it Interestingly, the fit to the combination of CMB and BAO is as good as for $\Lambda$CDM,} with the worsening of the Planck $\chi^2$ compensated by the improvement in the DESI fit. This can be seen from Table~\ref{tab:params} that lists the mean values and standard deviations in $b_{\rm pmf}$, $H_0$, $S_8$ and $\Omega_m$ obtained from PL+DESI and PL+DESI+PP+$M_b$, along with the best fit PL, DESI and PL+DESI $\chi^2$ values, comparing them to those from the $\Lambda$CDM PL+DESI fit.

Fig.~\ref{fig:pmf4}, along with Table~\ref{tab:params}, shows that the mild to moderate preference for the PMF is accompanied by a partial relief for both the $H_0$ and the $S_8$ tensions. It worsens the $\Omega_m$ tension with the SN data, which is the case for any modified recombination solution to the Hubble tension~\cite{Lee:2022gzh,Lynch:2024hzh,Poulin:2024ken,Mirpoorian:2024fka}. There are good reasons to expect the $\Omega_m$ problem to be entirely separate from the Hubble tension and any physics at recombination~\cite{Mirpoorian:2025rfp,Afroz:2025iwo}. The tension is between {\it uncalibrated} SN and BAO, and would remain in any model with a $\Lambda$CDM expansion history at $z<2$. We note that the $\Omega_m$ tension could be interpreted as evidence for dynamical dark energy~\cite{DESI:2024mwx} which, however, favours an even lower $H_0$. 

As previously noted in~\cite{Thiele:2021okz,Rashkovetskyi:2021rwg,Galli:2021mxk}, the CMB temperature and polarization spectra on very small angular scales, deep in the Silk damping tail, could be a powerful discriminant of the PMF proposal and other modified recombination models. With this in mind, we perform additional fits to the publicly available high-resolution CMB spectra from ACT~\cite{ACT:2020gnv} and SPT~\cite{SPT-3G:2022hvq}. Fig.~\ref{fig:actspt} shows the posteriors for $b_{\rm pmf}$ and $H_0$ from fits to Planck+DESI+ACT and Planck+DESI+SPT, with and without the ``SH0ES prior''. The corresponding $b_{\rm pmf}$ values and improvements in $\chi^2$ are provided in Table~\ref{tab:deltachi2}, while the full parameter tables are provided in the Appendix. While both ACT and SPT generally constrain $b_{\rm pmf}$ to smaller values, given the high sensitivity of the small-scale CMB anisotropies to the minute details of the ionization history~\cite{Mirpoorian:2024fka}, it is remarkable that the MHD-derived $x_e(z)$ still provide a good fit.

Fig.~\ref{fig:residuals} shows the CMB spectra residuals, {\it i.e.} the relative differences in $C_\ell^{TT}$, $C_\ell^{TE}$ and $C_\ell^{EE}$, for the best-fit PL+DESI+PP+$M_b$ $b\Lambda$CDM model relative to the PL+DESI best-fit $\Lambda$CDM model, along with the ACT and SPT data. The differences are at a few-percent level that will be well-within the constraining power of future CMB datasets from SPT-3G Ext-10k \cite{SPT-3G:2024qkd} and the Simons Observatory~\cite{Ade:2018sbj}.

\section{Discussion}

Using the most comprehensive evaluations of recombination in the presence of PMFs to date, our analysis shows that PMF-accelerated recombination is moderately favored by the combination of current (as of early March 2025) CMB, BAO, and SN observations. Even when fitting  $b\Lambda$CDM models to Planck data alone, we find a mild preference for  $b_{\rm pmf}$, having only one additional theoretical parameter, the final magnetic field strength at the onset of structure formation (around redshift  $z \approx 10$). The preference for  $b\Lambda$CDM strengthens when Planck data is combined with BAO measurements from DESI, ultimately reaching a  $\sim 3\sigma$  significance level with the inclusion of SH0ES-calibrated SN data. However, as described in the Methods section, there remain theoretical uncertainties due to the limited MHD simulation volumes.

While the goodness-of-fit to Planck data varies depending on the dataset combination, the fit to DESI data improves considerably. Notably, the PMF-modified recombination history, derived from MHD simulations with no free parameters beyond the PMF strength, provides an acceptable fit to Planck data. This is particularly intriguing given that CMB data is highly sensitive to changes in  $x_e(z)$, as demonstrated in the model-independent analysis of ~\cite{Mirpoorian:2024fka}, where fewer than 0.1\% of trial histories produced a good fit to Planck.

The origin of magnetic fields in galaxies, clusters, voids, and ridges remains an open question. Galactic and cluster magnetic fields may arise from seed fields amplified by small-scale dynamos and dispersed by supernova-driven outflows, though the details depend on the specific outflow model ~\cite{Marinacci:2017wew}. Alternatively, they may have emerged from primordial processes. A minimum field strength of $b_{\rm pmf} \approx 5$ pG would be required for cluster and galactic magnetic fields to have a purely primordial origin ~\cite{Banerjee:2003xk}. Intriguingly, this value closely aligns with that favored by cosmological data.

How can we firmly establish the presence of a PMF in the Universe? A promising avenue lies in future high-precision observations of the Silk damping tail of CMB anisotropies~\cite{Thiele:2021okz,Rashkovetskyi:2021rwg,Galli:2021mxk}. As Fig.~\ref{fig:residuals} suggests, there are likely differences between $\Lambda$CDM and $b\Lambda$CDM at high multipoles, which could ultimately lead to the exclusion of one of these models. Additional constraints may come from future direct measurements of $H_0$, including distance-ladder-based observations, as well as BAO surveys, which could further distinguish between the two models. Another promising approach involves upcoming  $\gamma$-ray observations of nearby blazars, which have the potential to place a lower limit of  $\sim 1$ pG on void magnetic fields~\cite{Korochkin:2020pvg}, a value remarkably close to those favored in this work. If PMFs originate from inflation, their effects on small-scale structure formation might be detectable in the Lyman-$\alpha$ forest \cite{Pavicevic:2025gqi}. Additionally, gravitational waves sourced by PMFs in the early Universe could provide another observational signature~\cite{RoperPol:2022iel}. Looking further ahead, more futuristic prospects include detecting CDM mini-halos induced by baryon clumping~\cite{Ralegankar:2023pyx} and the contribution of cosmological recombination radiation (CRR) to CMB spectral distortions~\cite{Lucca:2023cdl}.

Clearly, a firm detection of a PMF will require a multi-messenger approach. If achieved, it would provide invaluable insights into the early Universe evolution and potential extensions of the standard model of particle physics.

New data releases since the submission of our paper included the ACT DR6 \cite{AtacamaCosmologyTelescope:2025blo,AtacamaCosmologyTelescope:2025nti} and SPT-3G \cite{SPT-3G:2025bzu} CMB spectra, and DESI DR6 BAO \cite{DESI:2025zgx}, which improve significantly over the datasets used in this work. The analyses in \cite{AtacamaCosmologyTelescope:2025nti} and \cite{SPT-3G:2025bzu} included constraints on modified recombination based on the {\tt ModRec} model~\cite{Lynch:2024gmp}. They find a mild ($\sim 2\sigma$) preference for earlier recombination when CMB is combined with DESI DR2 BAO.

\section{Methods}

\subsection{Ionization histories from MHD simulations}
\label{sec:MHD}

For the computation of the average ionized fraction $\bar{x}_e(z)$ during recombination in the presence of PMFs, we use a suite of detailed MHD simulations. The magnetohydrodynamic evolution of the magnetized baryon fluid is computed with a modified version of the publicly available code {\tt ENZO}~\cite{ENZO:2013hhu}. This is coupled to the evolution of the local electron density computed using a private cosmic recombination code. The code mimics the code {\tt RECFAST}~\cite{Seager:1999bc,Wong:2007ym} and reproduces its results in a homogeneous, non-magnetized Universe to better than $0.5\%$ accuracy. The evolution of the baryon gas takes into account the drag force exerted by the CMB photons on moving baryons. The local speed of sound and the drag force are computed from the local ionized fraction. The full set of equations used in the MHD simulations are given in Section II of~\cite{Jedamzik:2023rfd}.

It is well known (cf. also Section IIIB of \cite{Jedamzik:2023rfd}) that the efficiency of recombination depends on the evolution of the Lyman-$\alpha$ photon occupation number. For one true recombination into the ground state of the hydrogen atom, a Lyman-$\alpha$ photon has to be lost. This may happen either via conversion into a two-photon state or by redshifting. In Section IV of~\cite{Jedamzik:2023rfd}, this process was investigated in detail in an inhomogeneous Universe. It was found that local Lyman-$\alpha$ loss rates are changed in a clumpy medium due to Lyman-$\alpha$ photon mixing between high- and low-density regions. The private recombination code thus also accounts for the mixing of Lyman-$\alpha$ photons, which is important at small coherence scales. In our simulation, we assume 100\% mixing of Lyman-$\alpha$ photons between different regions, inspired by the results of detailed Monte Carlo simulations of Lyman-$\alpha$ photon transport presented in Section IIIB of~\cite{Jedamzik:2023rfd}. In principle, additional Lyman-$\alpha$ loss may also occur due to peculiar motions in an inhomogeneous Universe. This has been investigated in Section IIIB of \cite{Jedamzik:2023rfd} and was found to influence $\bar{x}_e$ at less than the 1\% level. This process has therefore been neglected in our calculations.

\begin{table}[!htbp]
\centering
\begin{tabular}{c|c|c}
$b_{\rm pmf}$ & Run A & Run B \\ 
              & $\Delta L$ and $b_{ini}$ & $\Delta L$ and $b_{ini}$ \\
\hline \hline
$4.45 \times 10^{-3}$ nG & 0.5 - 4 kpc& 0.09375 -0.75 kpc \\
                         & $3.02 \times 10^{-2}$ nG & $1.98$ nG \\
$9.25 \times 10^{-3}$ nG & 1 - 8 kpc & 0.1875 - 1.5 kpc  \\
                         & $6.04 \times 10^{-2}$ nG & $3.96$ nG \\ 
$1.93 \times 10^{-2}$ nG & 2 - 16 kpc & 0.375 - 3 kpc \\
                         & $1.21 \times 10^{-1}$ nG & $7.92$ nG \\
$3.66 \times 10^{-2}$ nG & 4 - 32 kpc & 0.75 - 6 kpc \\
                         & $2.42 \times 10^{-1}$ nG & $16.84$ nG \\
 \hline \\
\end{tabular}
\label{tab:scales}
\caption{\label{tab:scales} 
For each of the four utilized normalizations of the Batchelor spectrum, two independent simulations are performed: (A) a simulation to determine the final field strengths, and (B) a simulation which determines the impact on recombination (cf. Section VB of \cite{Jedamzik:2023rfd} for details). The first column shows the total final comoving magnetic field strength at $z = 10$. The second and third columns show the length ranges and initial total comoving rms magnetic field strengths simulated for Run A and Run B, respectively. Note that the smallest modes are resolved by 32 zones.} 
\end{table}

\begin{table}[!htpb]
\centering
\begin{tabular}{c|c|c}
$b_{\rm pmf}$ & $b_{\rm pmf}^{\rm rec}$ & $B_{1 {\rm Mpc}}$ \\
\hline \hline
$4.45 \times 10^{-3}$ nG & $5.45 \times 10^{-2}$ nG & $1.68 \times 10^{-10}$ nG \\
$9.25 \times 10^{-3}$ nG & $1.04 \times 10^{-1}$ nG & $1.90 \times 10^{-9}$ nG \\
$1.93 \times 10^{-2}$ nG & $1.89 \times 10^{-1}$ nG & $2.15 \times 10^{-8}$ nG \\
$3.66 \times 10^{-2}$ nG & $3.65 \times 10^{-1}$ nG & $2.44 \times 10^{-7}$ nG \\
 \hline \\
\end{tabular}
\label{tab:brec}
\caption{\label{tab:brec} 
The simulations assume a non-helical magnetic field with Batchelor spectrum.
Total comoving magnetic field strengths at $z=10$, $b_{\rm pmf}$, for the MHD simulations used in this paper, and the corresponding total comoving field strengths $b_{\rm pmf}^{\rm rec}$ evaluated at the peak of the visibility function ($z\approx 1090$). Also shown are the values of $B_{1 {\rm Mpc}}$ commonly used in the literature on CMB anisotropy constraints on PMF.}
\end{table} 

The simulations also take into account the hydrodynamic heating of baryons by the dissipation of magnetic fields. Ambipolar diffusion, which is of some importance at lower redshifts $z \sim 100 - 300$, has not been included. Changing the low-redshift $\bar{x}_e$ ``by hand,'' we have observed that such changes do not impact the conclusions of our analysis.

The relatively small simulations of $256^3$ zones are computationally demanding and require about $2$ CPU-years. The smallness of the simulation volume results in significant realization variance. We discuss this issue in the next section. A detailed analysis presented in Section VB of \cite{Jedamzik:2023rfd} makes sure that the most relevant modes influencing $\bar{x}_e(z)$ are simulated. Some residual dependence on unresolved UV modes may still remain. In essence, for each magnetic field strength we include all the modes equivalent to those shown in Fig. 10 of ~\cite{Jedamzik:2023rfd}, {\it i.e.} modes which produce their peak in baryon clumping between $z \approx 800$ and $z\approx 3.3\times 10^4$. Here, the peak redshift was computed via the relation $v_A^2/L^2 = \alpha H$ \cite{Banerjee:2004df}, where $v_A$, $L$, $\alpha$, and $H$ are the Alfven velocity, the coherence scale, drag constant and the Hubble parameter, respectively. This relation was numerically confirmed in \cite{Jedamzik:2023rfd} (cf. Fig. 7). A test if even smaller UV modes impact the results was not possible due to numerical resource limitations. It is noted that due to the smallness of our simulations it is impossible to determine if an inverse cascade and the conservation of the Hosking integral also applies in the viscous regime \cite{Hosking:2020wom}.

All our simulations use a non-helical magnetic field with a Batchelor spectrum, and four different normalizations of the spectrum were simulated. For the cosmological parameters in the simulations, we used a baryon density $\omega_b=0.0224$, a CDM density $\omega_c=0.12$, a Hubble constant $h=0.67$, and a helium fraction $Y_\mathrm{P}=0.24$,  close to those of the best-fit Planck $\Lambda$CDM model. For changes in the cosmological parameters, the reader is referred to the next section. Final field strengths at $z = 10$, $b_{\rm pmf}$, are computed by independent simulations, taking into account those modes which dominate the final field. We give the simulated length scale ranges and initial magnetic field strengths for both simulations, the one which determines the final field strength and the one which computes the impact on recombination, in Table~\ref{tab:scales}. The corresponding field strengths at recombination ($z \approx 1090$) are approximately a factor of ten larger and are provided in Table~\ref{tab:brec}. This is due to the additional damping that takes place shortly after recombination, when the plasma transitions from a highly viscous to a turbulent state. Between $z \approx 100$ and $z = 0$, the damping is logarithmic~\cite{Banerjee:2004df}, which means the present-day PMF is about 17\% smaller than $b_{\rm pmf}$, but still comparable to it. Constraints on PMFs from CMB anisotropies are often presented in terms of the field strength smoothed over a 1 Mpc size region, $B_{1 \, {\rm Mpc}}$, and the magnetic spectrum index $n_B$~\cite{Paoletti:2010rx,Planck:2015zrl,Zucca:2016iur}. For a Batchelor spectrum, $n_B = 2$ (in the notation where the scale-invariant spectrum has $n_B = -3$), making $B_{1 \, {\rm Mpc}}$ negligible compared to $b_{\rm pmf}^{\rm rec}$.

\subsection{Obtaining ionization histories for general values of PMF strengths and cosmological parameters}
\label{sec:ionization}

\begin{figure*}[htbp!]
\includegraphics[width=0.49\textwidth]{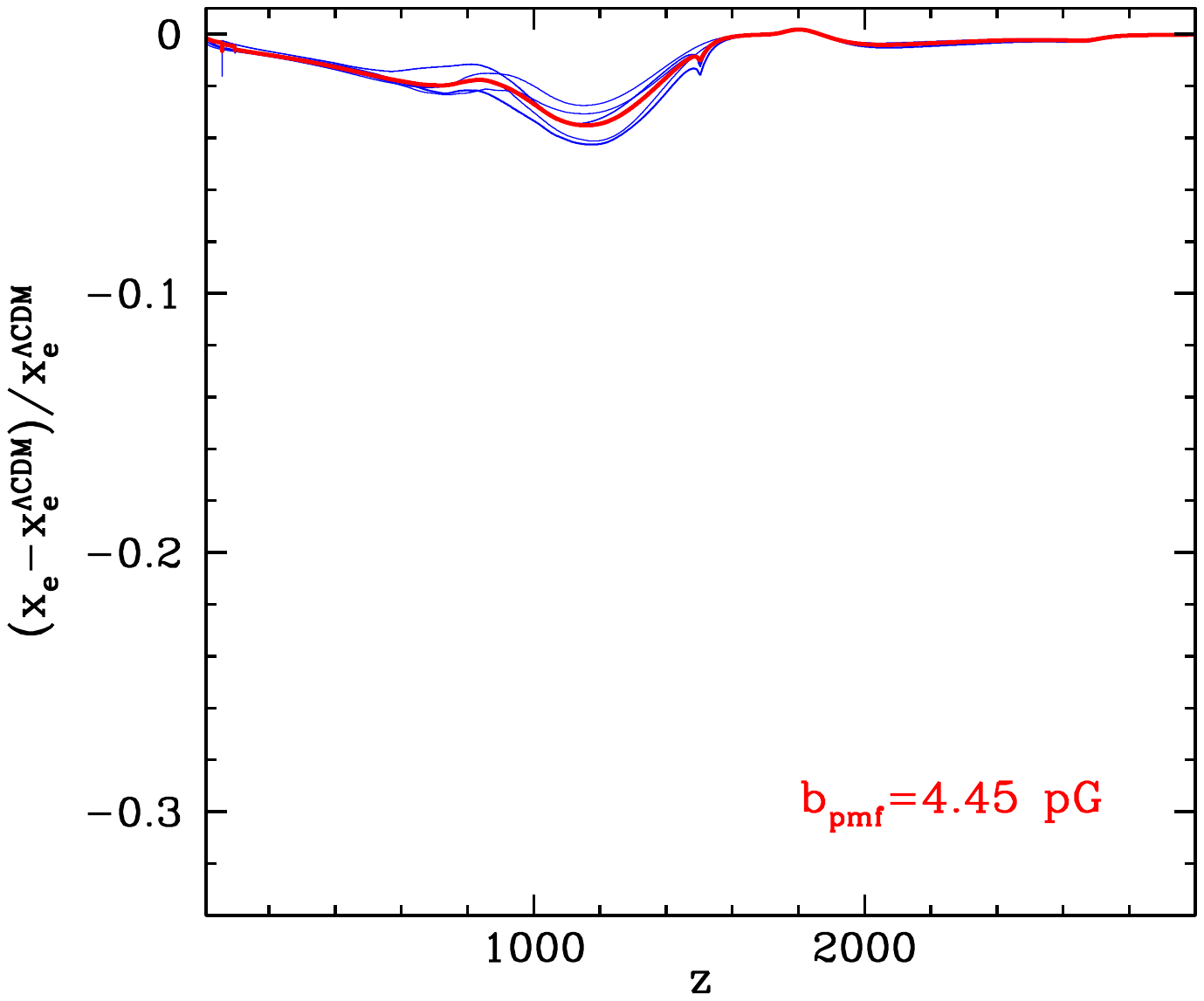}
\includegraphics[width=0.49\textwidth]{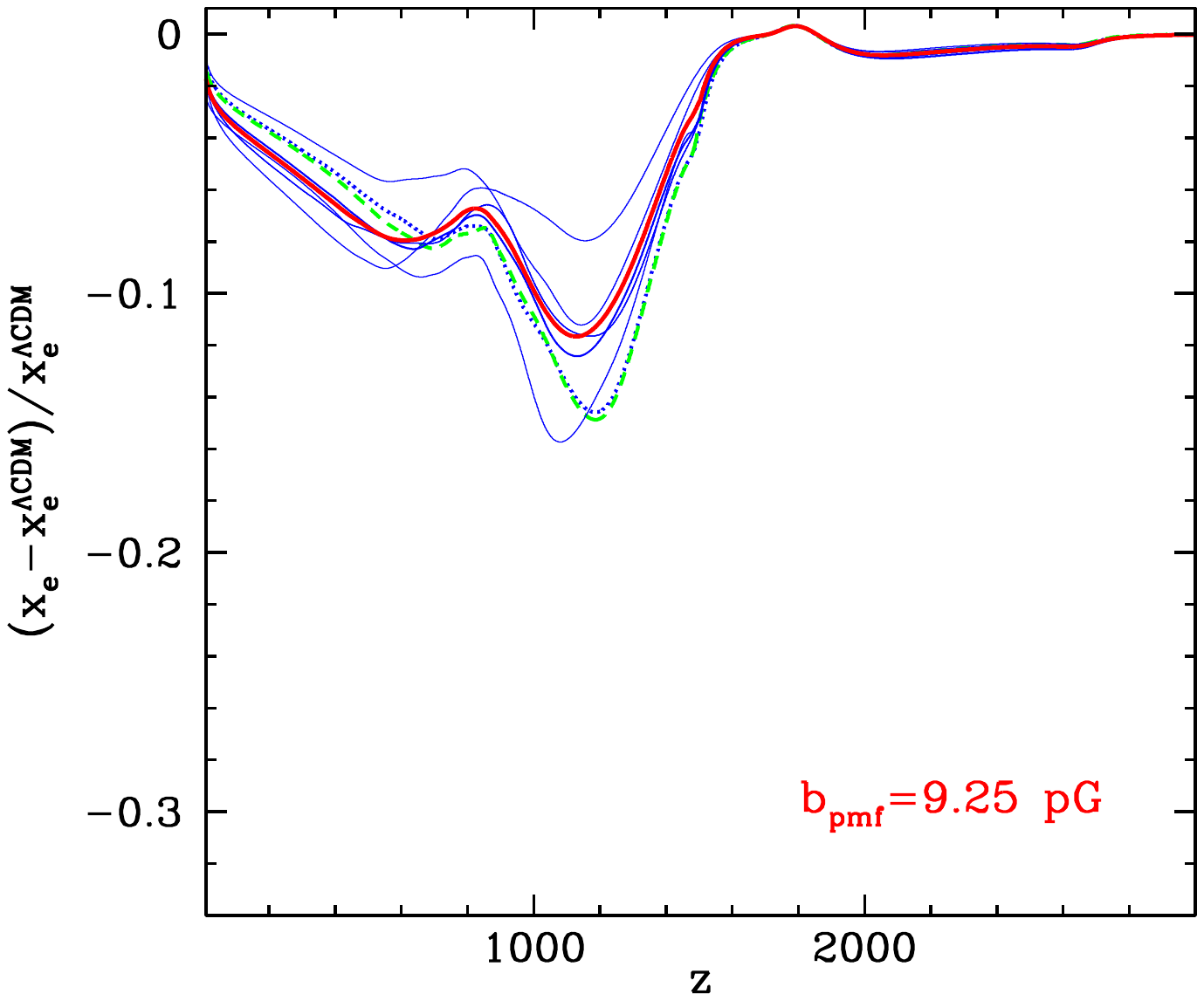}
\includegraphics[width=0.49\textwidth]{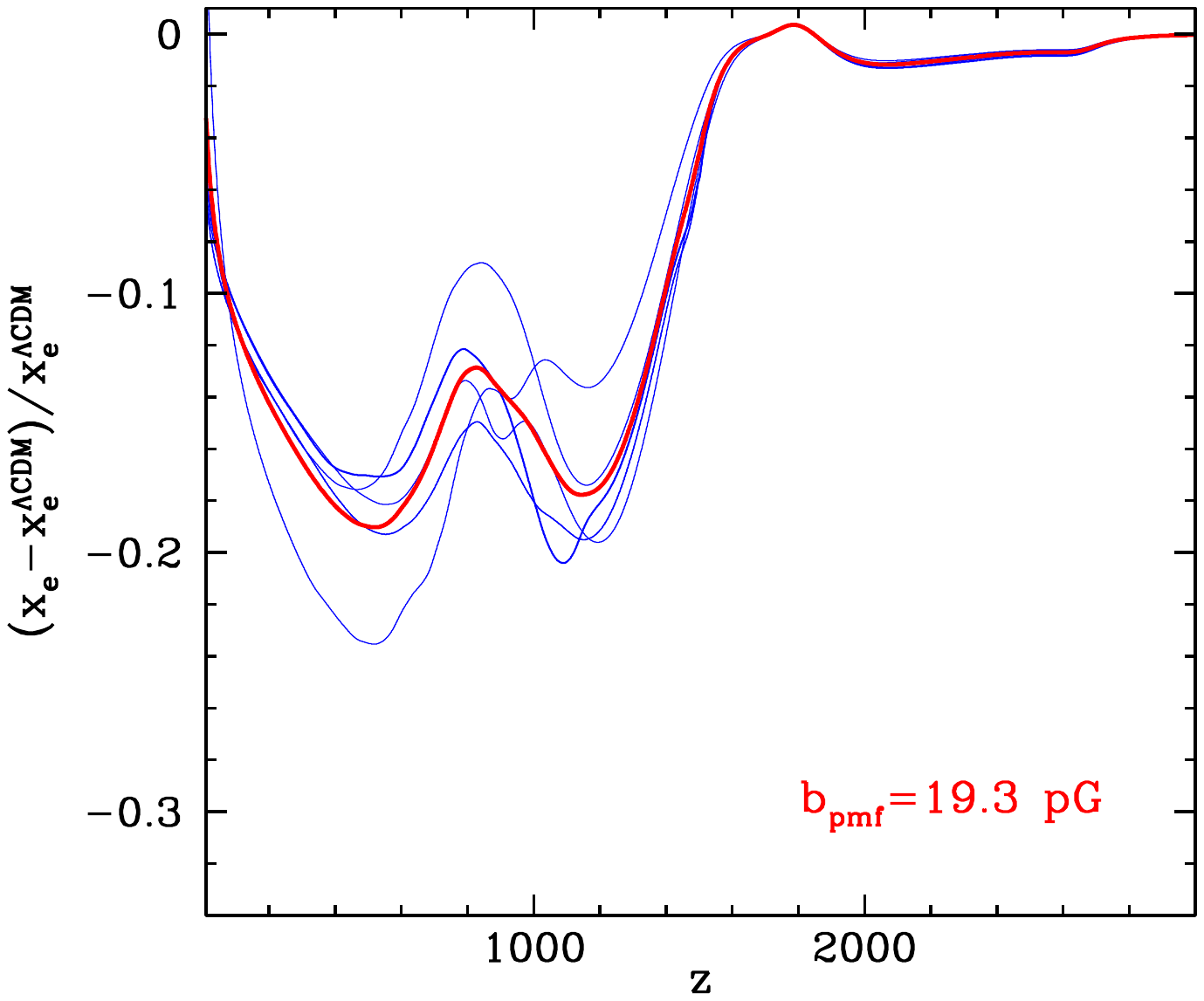}
\includegraphics[width=0.49\textwidth]{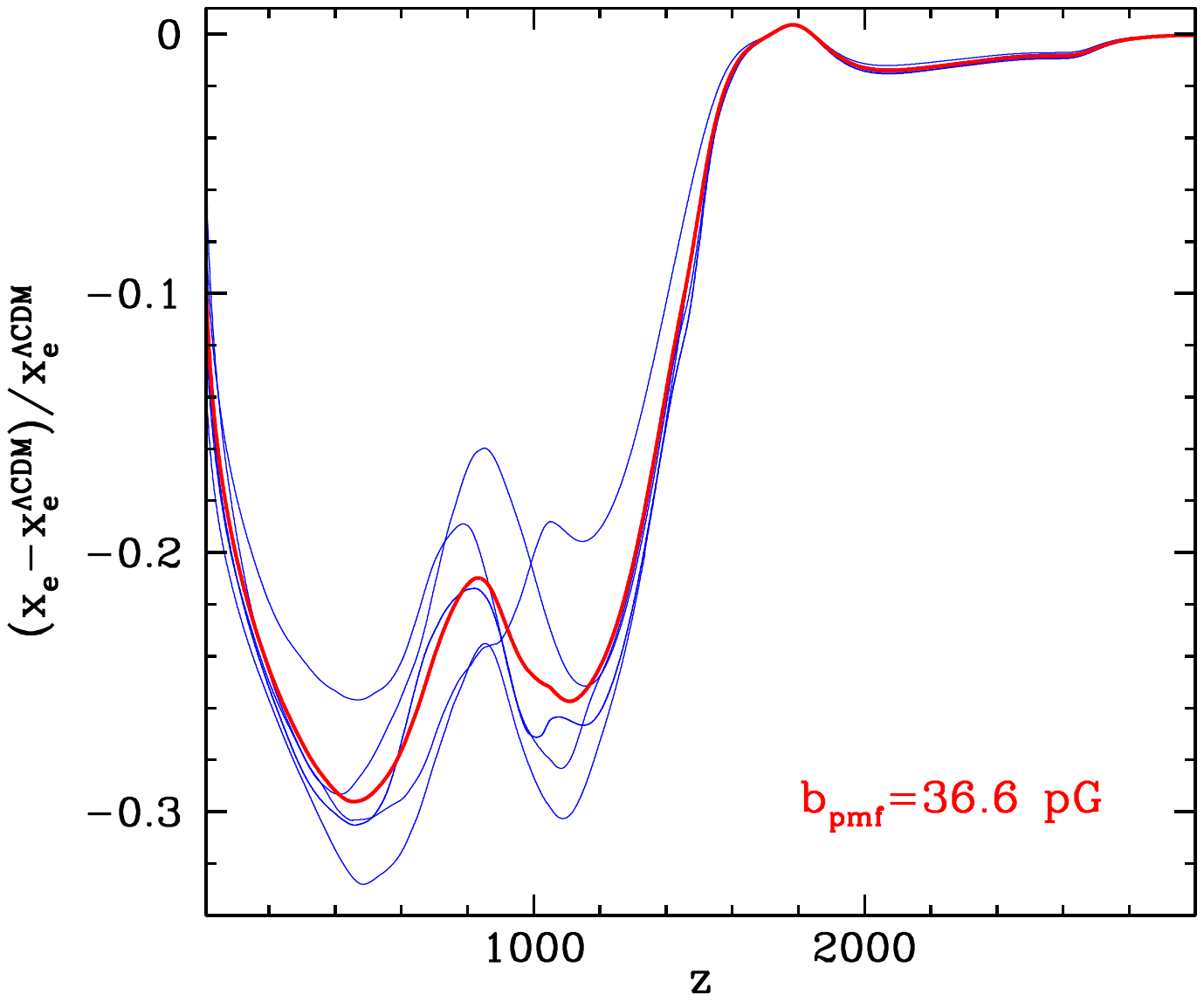}
\caption{\label{fig:xe} 
Relative differences in the ionization histories $x_e(z)$ with respect to the reference $\Lambda$CDM model obtained from MHD simulations with total $z=10$ comoving magnetic field strength of 4.45 pG (top-left), 9.25 pG (top-right), 19.3 pG (bottom-left) and 36.6 pG (bottom-right). The present-day field strength values are approximately 17\% lower, while the corresponding values at recombination are given in Table \ref{tab:brec} and are a factor of 10 larger. The blue lines are the results from the 5 realizations at each $b_{\rm pmf}$ with the mean ionization history shown in red. The top-right panel includes a 6th realization evaluated at the fiducial cosmological parameters (blue dot) and an alternative set of parameters (green dash), as described in Sec~4.2, demonstrating that changes in $\Delta_e(z)$ due to moderate changes in cosmological parameters are minor compared to the sample variance.}
\end{figure*}

The MHD simulations provide us with the difference in $x_e(z)$ relative to the fiducial $\Lambda$CDM model ionization history between $z=4500$ and $z=10$ for four values of the comoving PMF strengths at $z=10$: $b_{\rm pmf} = 4.45$ pico-Gauss (pG), $9.25$ pG, $19.3$ pG and $36.6$ pG. The set of fiducial cosmological parameters, $\bf{\Omega}^{\rm fid}$, has $\omega_b=0.0224$, $\omega_c=0.12$, $h=0.67$, and $Y_\mathrm{P}=0.24$. The relative difference is defined as
\be
\Delta_e(z, {\bf \Omega}^{\rm fid}) = { x_e^{\rm pmf}(z, {\bf \Omega}^{\rm fid} ) - x_e^{\rm \Lambda CDM}(z, {\bf \Omega}^{\rm fid}) \over x_e^{\rm \Lambda CDM}(z, {\bf \Omega}^{\rm fid} ) } .
\ee
We then use linear interpolation to obtain $\Delta_e(z, {\bf \Omega}^{\rm fid})$ for other values of $b_{\rm pmf}$, with $\Delta_e(z, {\bf \Omega}^{\rm fid}) = 0$ for $b_{\rm pmf}=0$ by definition. We assume that the dependence of $x_e^{\rm pmf}(z)$ on the cosmological parameters is largely the same as in $\Lambda$CDM, implying that $\Delta_e(z)$ is effectively independent of cosmological parameters for models that are reasonably close to $\Lambda$CDM. To illustrate that this is a good assumption, the top-right panel in Fig.~\ref{fig:xe} shows an additional realization of $\Delta_e(z)$ evaluated at two different cosmological parameters that are approximately $2\sigma$ apart in the Planck $\Lambda$CDM cosmology. One of them is $\bf{\Omega}^{\rm fid}$, while the other has $\omega_b=0.02237$, $\omega_c=0.1228$ and $h=0.682$. On can see that the differences between the two are much smaller than the sample variance. With that assumption, we obtain $x_e^{\rm pmf}(z)$ for a given vector of cosmological parameters $\bf{\Omega}$ as
\be
x_e^{\rm pmf}(z, {\bf \Omega}, b_{\rm pmf}) = x_e^{\rm \Lambda CDM}(z, {\bf \Omega}) [1+ \Delta_e(z, \bf{\Omega}^{\rm fid},b_{\rm pmf})],
\ee
and implement it in {\tt RECFAST} inside {\tt CAMB} that is used with {\tt Cobaya}~\cite{Torrado:2020dgo} to constrain the parameters.

\subsection{Using randomly generated ensemble of ionization histories instead of the mean $x_e(z)$}
\label{sec:ensemble}

For the main results of this paper, we used the mean ionization histories obtained by averaging over 5 MHD realizations. Doing so does not account for the large variance in $\Delta_e(z)$ from realization to realization clearly seen in Fig.~\ref{fig:xe}. Strictly speaking, the constraints we derive on $b_{\rm pmf}$ and the cosmological parameters should account for the theoretical uncertainty in $\Delta_e(z)$, which in our case is dominated by the sample variance of the MHD-derived results.  To investigate the impact of accounting for the sample variance in $\Delta_e(z)$, we ran MCMC chains for one data combination, Planck+DESI+PP+$M_b$, using ionization histories randomly generated from a Gaussian distribution with the mean and covariance derived from the 5 MHD realizations at each PMF strength.

\begin{figure}[htbp!]
\includegraphics[width=0.48\textwidth]{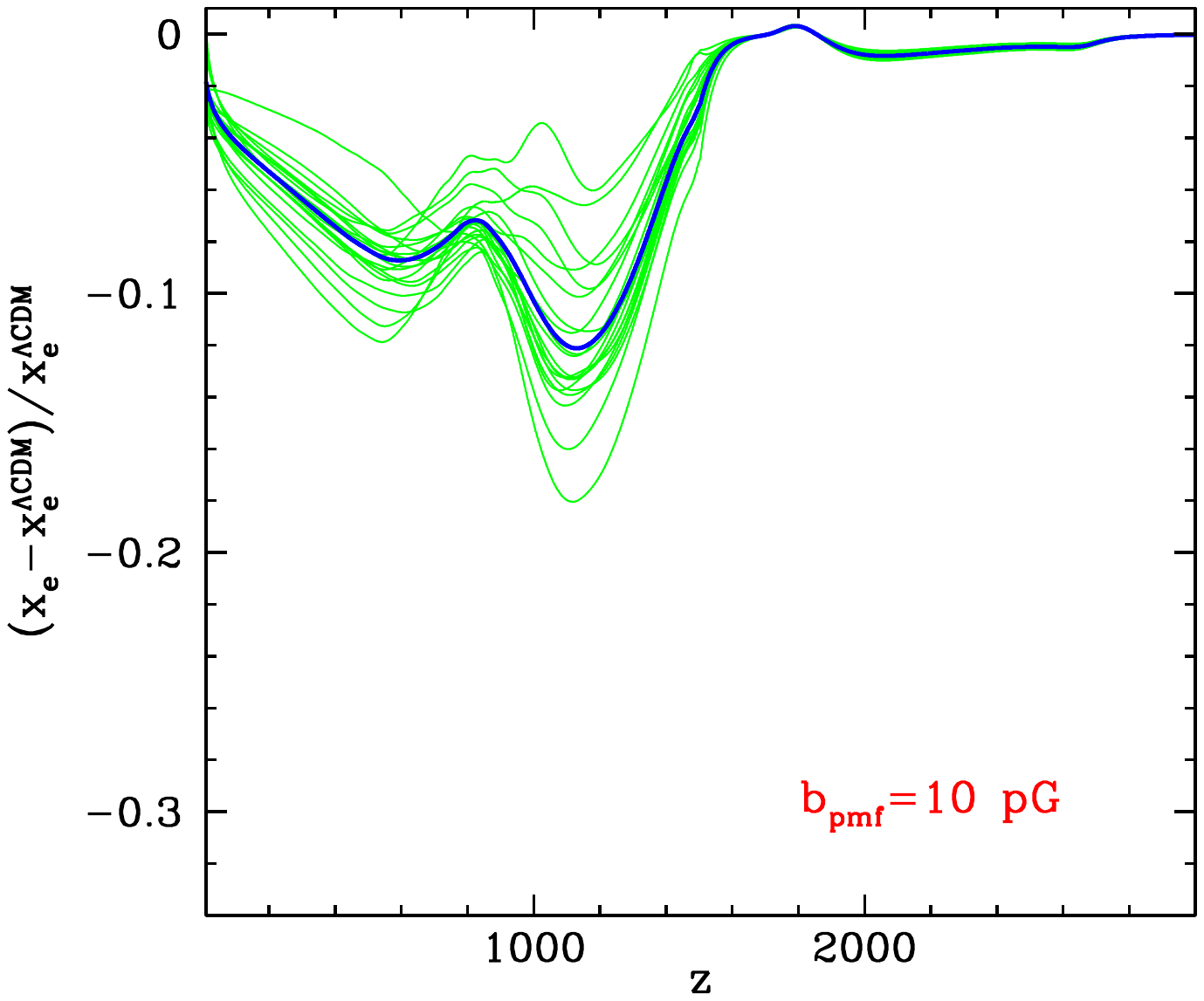}
\caption{\label{fig:random} $\Delta_e(z)$ generated randomly from a Gaussian distribution a mean and covariance corresponding to $b_{\rm pmf}=10$ pG. The figure shows 20 random histories in green and the mean in blue.}
\end{figure}

The ensemble of realizations was generated as follows. Let vector ${\bf x} = \{x_i\}$ represent $\Delta_e(z_i)$ at a set of discrete values of redshift $z_i$. For a given $b_{\rm pmf}$, we have 5 realizations of ${\bf x}$ from which we can compute the mean, $\langle {\bf x} \rangle$ and the covariance $C \equiv \langle ({\bf x}-\langle {\bf x} \rangle)({\bf x}-\langle {\bf x} \rangle)^T\rangle$. 

To generate a random vector ${\bf y}$ from a Gaussian distribution of covariance $C$ and mean $\langle {\bf x} \rangle$, we first use the Singlular Value Decomposition (SVD) method to decompose the covariance into $C = UWV^T$, where $U$ and $V$ are orthogonal matrices and $W$ is a diagonal matrix of singular values of $C$. For a symmetrical matrix $C$, such as our covariance, $U=V$ and the elements of $W$ are the eigenvalues of $C$. Next, we consider a vector ${\bf g}$ drawn from a Gaussian distribution of zero mean and unit covariance, {\it i.e.} $\langle {\bf g} \rangle = 0$ and $\langle {\bf g}{\bf g}^T \rangle = \mathbb{I}$, and build our random vector as ${\bf y} = \langle {\bf x} \rangle + L {\bf g}$, where $L=U \sqrt{W}$. One can readily check that $\langle {\bf y} \rangle = \langle {\bf x} \rangle$, and $\langle ({\bf y}-\langle {\bf y} \rangle)({\bf y}-\langle {\bf y} \rangle)^T\rangle = \langle L{\bf g} {\bf g}^T L^T \rangle =L  \langle {\bf g} {\bf g}^T\rangle L^T = U W U^T = C$. Thus, the random vector ${\bf y}$ has the desired mean and covariance. In practice, only a handful of eigenvalues have non-negligible values, and thus only a small number of eigenvectors of $C$ (or rows of matrix $L$), corresponding to the largest eigenvalues, need to be stored to accurately generate ${\bf y}$. 

The mean and covariance for a given $b_{\rm pmf}$ are obtained by linearly interpolating over the means and covariances at 4 $b_{\rm pmf}$ values available from MHD simulations. Fig.~\ref{fig:random} shows a collection of randomly generated $\Delta_e(z)$ corresponding to $b_{\rm pmf}=10$ pG.

\begin{figure}[htbp!]
\includegraphics[width=0.48\textwidth]{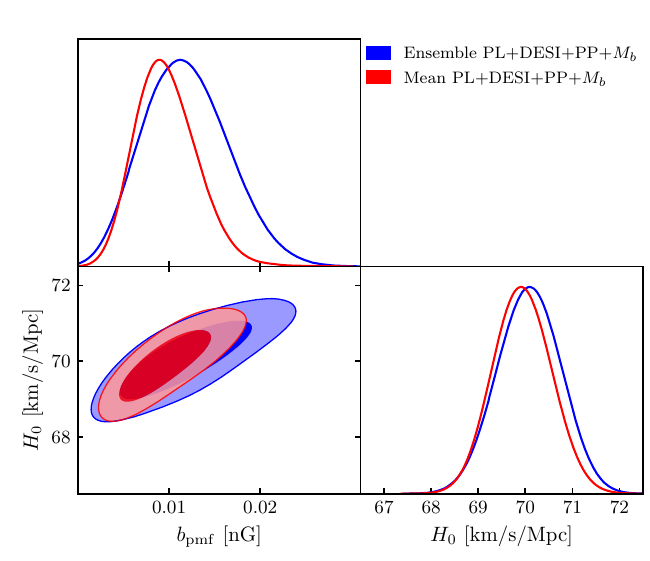}
\caption{\label{fig:ensemble} The marginalized $68$\% and $95$\% CL comparing using the mean vs randomly generated $x_e(z)$ when fitting to PL+DESI+PP+$M_b$.}
\end{figure}

Fig.~\ref{fig:ensemble} compares posteriors of $b_{\rm pmf}$ and $H_0$ obtained by fitting an ensemble of randomly generated $x_e(z)$ using the above method {\it vs} using the mean $x_e(z)$ as in the main text of the paper for PL+DESI+PP+$M_b$. As expected, the posteriors are generally wider, especially for  $b_{\rm pmf}$, but the differences in constraints on cosmological parameters, including $H_0$, are minor. We obtain $b_{\rm pmf} = 0.0119^{+0.0040}_{-0.0050}$ nG from the ensemble-based MCMC, compared to $b_{\rm pmf} = 0.0096^{+0.0029}_{-0.0036}$ nG we obtained from the same data combination while using the mean $x_e(z)$, corresponding to a 38\% increase in uncertainty, and roughly the same statistical significance of the preference for $b_{\rm pmf}$. The corresponding values of the Hubble constant were $H_0 = 70.08 \pm 0.63$ and $H_0 = 69.93^{+0.53}_{-0.66}$ km/s/Mpc for the ensemble-based and the mean-based runs, respectively. The convergence of the ensemble-based MCMC runs is very slow, which is why most of our results are based on using the mean ionization histories. As more MHD runs, or runs using larger simulation boxes become available, the uncertainty around the mean will decrease and the difference between the two methods should disappear, and the MCMC convergence should improve.

\subsection{Datasets and parameter priors}
\label{sec:datasets}

Our datasets included the $\ell > 30$ CMB temperature (TT), polarization (EE) and cross-correlation (TE) angular spectra, $C^{\rm TT}_\ell$, $C^{\rm EE}_\ell$ and $C^{\rm TE}_\ell$, as implemented in the Planck PR4 CamSpec ({\tt NPIPE}) likelihood \cite{Rosenberg:2022sdy}, the Planck 2018 baseline $\ell<30$ TT and EE likelihoods~\cite{Planck:2018vyg}, and the Planck PR4 CMB lensing likelihood~\cite{Carron:2022eyg}. In addition, we used the 12 BAO measurements at 7 effective redshifts from the DESI Year 1 release~\cite{DESI:2024mwx}, and the Pantheon+ (PP) SN dataset~\cite{Brout:2022vxf} calibrated using the brightness magnitude $M_b = -19.253 \pm 0.027$ from the cosmic distance ladder measurement by the SH0ES collaboration~\cite{Riess:2021jrx}. We also use the high-resolution CMB TT, TE, EE spectra from the 2022 release of the South Pole Telescope 3G (SPT-3G) experiment~\cite{SPT-3G:2022hvq} and from the 4th data release of the Atacama Cosmology Telescope experiment (ACT-DR4)~\cite{ACT:2020gnv}. 

We use {\tt Cobaya}~\cite{Torrado:2020dgo} to perform the MCMC analysis and derive the posterior distributions of the model parameters. The priors for the standard cosmological parameters are those commonly used in CMB analyses. Namely, they are sampled uniformly in the ranges given by $\log A \in [1.61, 3.91]$,  $n_s \in [0.8,1.2]$,  $\theta_{MC} \in [0.5,10]$, $\omega_\mathrm{b} \in [0.005, 0.1]$, $\omega_\mathrm{c} \in [0.001,0.99]$ and $\tau \in [0.01,0.8]$. In addition, we sample the magnetic field strength uniformly in the range $b_{\rm pmf} \in [0.0, 0.037{\rm nG}]$.

{\it Acknowledgments.}  We thank Andrei Frolov and Hamid Mirpoorian for valuable discussions and technical assistance. This research was enabled in part by support provided by the BC DRI Group and the Digital Research Alliance of Canada ({\tt alliancecan.ca}). KJ is supported in part by ANR grant COSMAG. LP is supported in part by the National Sciences and Engineering Research Council (NSERC) of Canada. TA is supported by the U.S. Department of Energy, Office of Science under contract number DE-AC02-76SF00515. For the purposes of open access, the authors have applied a CC BY public copyright license to any Author Accepted Manuscript arising from this submission, irrespective of the publication model under which the article is published.

\appendix
\section{Parameter tables}

\begin{table*}[!tbp]
\centering
\begin{tabular}{c|c|c|c|c|c|c}
 $\Lambda$CDM &  PL & DESI & DESI+$\theta_\mathrm{CMB}$ & PL+DESI & PL+DESI+PP & PL+DESI+PP+$M_b$   \\
\hline \hline
{\boldmath$\log(10^{10} A_\mathrm{s})$}  & $3.038\pm 0.014$ & - & - &  $3.045\pm 0.014$ & $3.043\pm 0.014$ & $3.052\pm 0.014            $  \\
{\boldmath$n_\mathrm{s}   $} & $0.9635\pm 0.0040$ & -&  - & $0.9671\pm 0.0036$ & $0.9663\pm 0.0036$ & $0.9702\pm 0.0035          $ \\
{\boldmath$100\theta_\mathrm{MC}$} & $1.04075\pm 0.00025$ & -&  - & $1.04092\pm 0.00024$ & $1.04089\pm 0.00024$ & $1.04111\pm 0.00024        $ \\
{\boldmath$\Omega_\mathrm{b} h^2$} & $0.02218\pm 0.00013$ & -&  - & $0.02228\pm 0.00013 $ & $0.02226\pm 0.00012$ & $0.02242\pm 0.00012        $ \\
{\boldmath$\Omega_\mathrm{c} h^2$} & $0.1197\pm 0.0011$ & -& -&  $0.11829\pm 0.00081$ & $0.11859\pm 0.00078$  & $0.11707\pm 0.00075        $ \\
{\boldmath$\tau_\mathrm{reio}$} & $0.0526\pm 0.0071$ &- & -& $0.0573\pm 0.0072$ & $0.0563\pm 0.0070$ & $0.0618\pm 0.0073          $ \\
$H_0 {\rm [km/s/Mpc]}$ & $67.22\pm 0.47$ &-&  -& $67.88\pm 0.37$ & $67.74\pm 0.35$  & $68.49\pm 0.34             $ \\
$\Omega_\mathrm{m}         $ & $0.3156\pm 0.0065$ & $0.295 \pm 0.014$ & $0.295 \pm 0.008$ & $0.3066\pm 0.0049$ & $0.3084\pm 0.0047$  & $0.2988\pm 0.0044          $ \\
$\Omega_\mathrm{m} h^2     $ & $0.14257\pm 0.00099$  &-& -&  $0.14122\pm 0.00077$ & $0.14150\pm 0.00075$ & $0.14014\pm 0.00072        $ \\
$\sigma_8$ & $0.8078\pm 0.0055$ & -& -&     $0.8066\pm 0.0056$ & $0.8069\pm 0.0056$  & $0.8058\pm 0.0058          $ \\
$S_8$ & $0.828\pm 0.011$ &-& -&       $0.8154\pm 0.0090 $ & $0.8181\pm 0.0089$  & $0.8042\pm 0.0086          $ \\
$r_\mathrm{drag} h \ {\rm [Mpc]}$ & $99.07\pm 0.81$ & $101.89 \pm 1.25$ & $101.89 \pm 1.01$ & $100.22\pm 0.63$ & $99.98\pm 0.60$  & $101.24\pm 0.59            $ \\
$r_\mathrm{drag}  \ {\rm [Mpc]}$ & $147.38 \pm 0.24$ &-& -& $147.65\pm 0.21$ & $147.60\pm 0.21$  & $147.82\pm 0.21            $  \\
$z_{\rm{drag}}$ &  $1059.48\pm 0.28           $ & -& -&    $1059.61\pm 0.28           $ & $1059.58\pm 0.27           $ & $1059.84\pm 0.27           $ \\
$r_\mathrm{s} \ {\rm [Mpc]}$ & $144.65\pm 0.24            $ & - & -&   $144.95\pm 0.20            $  & $144.89\pm 0.19            $ & $145.16\pm 0.19            $ \\
$z_\mathrm{s}$ & $1090.14\pm 0.23           $ & -& - &   $1089.88\pm 0.20           $  & $1089.93\pm 0.19           $ & $1089.60\pm 0.18           $ \\
$M_b$ & - & - & - & - & - & $-19.406 \pm 0.010$ \\
\hline
$\chi^2_{\rm highl}         $ & $10543.7$ &-&  -& $10545.6$ & $10546.7$ & $10549.3$  \\
$\chi^2_{\rm lensing}      $ & $8.51$ &-&  -& $8.54$ & $8.79$ & $9.28$ \\
$\chi^2_{\rm lowlTT}         $ & $23.72$ &-&  -& $22.75$ & $22.66$ & $22.60$ \\
$\chi^2_{\rm lowlEE}       $ & $395.72$ &-& -& $397.02$ & $396.39$  & $398.75$ \\
$\chi^2_{\rm PL}             $ & $10971.7$ &-&  -& $10973.9$ & $10974.5$ & $10979.9$ \\
$\chi^2_{\rm DESI}         $ & - & $14.71$ & $12.74$ & $16.55$ & $15.55$ & $13.20$ \\
$\chi^2_{\rm PP}         $ & - &-& - & - & $1404.9$ & $1406.8$ \\
$\chi^2_{M_b}         $ & - &-& - & - & - & $32.12$ \\
$\chi^2_{\rm PL+DESI}         $ & -&  -& -& $10990.4$ & $10990.0$ & $10,993.1$  \\
$\chi^2_{\rm PL+DESI+PP}         $ & - & -& -& - & $12395.0$  & $12399.9$ \\
$\chi^2_{{\rm PL+DESI+PP}M_b}         $ & - & -& -& - & -  & $12432.0$ \\
 \hline \\
\end{tabular}
\caption{\label{tab:params_lcdm} 
The median values and the 68\% credible intervals of the {\bf primary} and selected derived parameters of the $\Lambda$CDM model fit to combinations of Planck (PL), DESI BAO, angular size of the CMB acoustic scale $\theta_\mathrm{CMB}$, and uncalibrated Pantheon+ (PP) SN, along with the best fit $\chi^2$ values.}
\end{table*}

\begin{table*}[!tbp]
\centering
\begin{tabular}{c|c|c|c|c}
  $b\Lambda$CDM &  PL & PL+DESI & PL+DESI+PP & PL+DESI+PP+$M_b$ \\
\hline \hline
{\boldmath$b_\mathrm{pmf} \ {\rm [nG]}$} & $0.0023^{+0.0048}_{-0.0022}$ & $0.0042^{+0.0021}_{-0.0028}$ &  $0.0038^{+0.0017}_{-0.0028}$ & $0.0096^{+0.0029}_{-0.0036}$ \\
{\boldmath$\log(10^{10} A_\mathrm{s})$}  & $3.041\pm 0.014            $  & $3.048\pm 0.014            $ & $3.047\pm 0.014            $ & $3.054\pm 0.014            $ \\
{\boldmath$n_\mathrm{s}   $} & $0.9668 \pm 0.0050$ & $0.9714\pm 0.0045          $ & $0.9700\pm 0.0044          $ & $0.9791\pm 0.0042          $ \\
{\boldmath$100\theta_\mathrm{MC}$} & $1.04201\pm 0.00124$ & $1.0429\pm 0.00149$ & $1.04260 \pm 0.00134$ & $1.0467\pm 0.0019          $ \\
{\boldmath$\Omega_\mathrm{b} h^2$} & $0.02233\pm 0.00018$ & $0.02248\pm 0.00017        $ & $0.02244\pm 0.00017        $ & $0.02279\pm 0.00016        $ \\
{\boldmath$\Omega_\mathrm{c} h^2$} &  $0.1198\pm 0.0010          $ & $0.11882\pm 0.00092        $ & $0.11906\pm 0.00088        $ & $0.1192\pm 0.0010          $ \\
{\boldmath$\tau_\mathrm{reio}$} & $0.0535\pm 0.0072          $ & $0.0573\pm 0.0071          $ & $0.0565\pm 0.0070          $ & $0.0589\pm 0.0073          $ \\
$H_0 {\rm [km/s/Mpc]}$ & $67.75^{+0.58}_{-0.74}$\  & $68.52^{+0.54}_{-0.62}$ & $68.29^{+0.49}_{-0.57}$ & $69.93^{+0.53}_{-0.66}$  \\
$\Omega_\mathrm{m}$ &  $0.3112\pm 0.0075          $ & $0.3024\pm 0.0055          $ & $0.3048\pm 0.0052          $ & $0.2917\pm 0.0048          $ \\
$\Omega_\mathrm{m} h^2$ &  $0.1428\pm 0.0010          $ & $0.14195\pm 0.00094        $ & $0.14214\pm 0.00089        $ & $0.1426\pm 0.0011          $ \\
$\sigma_8$ &  $0.8111 \pm 0.0063$ & $0.8122\pm 0.0068          $  & $0.8119\pm 0.0066          $ & $0.8210\pm 0.0074          $ \\
$S_8$ & $0.826\pm 0.011            $ & $0.8154\pm 0.0090          $ & $0.8184\pm 0.0088          $ & $0.8095\pm 0.0087          $ \\
$r_\mathrm{drag} h \ {\rm [Mpc]}$ & $99.62 \pm 0.97     $ & $100.77\pm 0.73            $ & $100.45\pm 0.69            $ & $102.24\pm 0.68            $ \\
$r_\mathrm{drag}  \ {\rm [Mpc]}$ & $147.05 \pm 0.37    $ & $147.06\pm 0.45  $ & $147.08 \pm 0.41 $ & $146.20\pm 0.53            $ \\
$z_{\rm{drag}}$ &  $1061.4 \pm 1.79  $ & $1062.6 \pm 2.14     $ & $1062.2\pm 1.93 $ & $1068.0\pm 2.7             $ \\
$r_\mathrm{s} \ {\rm [Mpc]}$ & $144.36 \pm 0.35   $ & $144.40\pm 0.43    $  & $144.41 \pm 0.39 $ & $143.59\pm 0.52            $ \\
$z_\mathrm{s}$ &  $1091.82 \pm 1.65  $ & $1092.7 \pm 2.11 $ & $1092.3 \pm 1.89$ & $1097.9\pm 2.8             $ \\
$M_b$ & - & - & - & $-19.364^{+0.016}_{-0.018} $ \\
\hline
$\chi^2_{\rm highl}$ &  $10544$  & $10546.1$ & $10544.8$ & $10551.2$ \\
$\chi^2_{\rm lensing}$ & $8.62$  & $8.56$ & $8.59$ & $9.32$\\
$\chi^2_{\rm lowlTT}$ & $21.66$ & $20.69$ & $20.71$ & $20.65$ \\
$\chi^2_{\rm lowlEE}$ &  $395.69$ & $396.21$ & $396.75$ & $396.46$\\
$\chi^2_{\rm PL}$ & $10970$ & $10971.6$ & $10970.9$ & $10977.6$\\
$\chi^2_{\rm DESI}$ & - & $14.07$ & $15.14$ & $12.91$ \\
$\chi^2_{\rm PP}$ & - & -  & $1405.13$ & $1408.34$ \\
$\chi^2_{M_b}$ & - &  - & - & $17.90$ \\
$\chi^2_{\rm PL+DESI}$ & - & $10985.7$ & $10986.04$ & $10990.51$ \\
$\chi^2_{\rm PL+DESI+PP}$ & - & $10985.7$ & $12391.17$ & $12398.85$ \\
$\chi^2_{{\rm PL+DESI+PP}M_b}$  & - & - & - & $12416.75$ \\
 \hline \\
\end{tabular}
\caption{\label{tab:params_pmf} 
The median values and the 68\% credible intervals of the {\bf primary} and selected derived parameters of the $b\Lambda$CDM model fit to combinations of Planck (PL), DESI BAO, uncalibrated Pantheon+ (PP) SN, PP calibrated using the SH0ES value of the intrinsic SN brightness (PP+$M_b$), along with the best fit $\chi^2$ values.}
\end{table*}

\begin{table*}[!tbp]
\centering
\begin{tabular}{c|c|c|c|c|c|c}
&  \multicolumn{2}{c|}{$\Lambda$CDM PL+DESI+} & \multicolumn{2}{c|}{$b\Lambda$CDM PL+DESI+} & \multicolumn{2}{c}{$b\Lambda$CDM PL+DESI+PP+$M_b$+}  \\
&ACT & SPT & ACT & SPT  & ACT & SPT \\
\hline \hline
{\boldmath$b_\mathrm{pmf} \ {\rm [nG]}$} & - & - &  $0.0030^{+0.0012}_{-0.0021}$ & $0.0036^{+0.0015}_{-0.0023}$ & $0.0064\pm 0.0021$ & $0.0074^{+0.0018}_{-0.0027}$ \\
{\boldmath$\log(10^{10} A_\mathrm{s})$}  & $3.051\pm 0.013$ & $3.042\pm 0.013$ & $3.055\pm 0.014            $ & $3.046\pm 0.014            $ & $3.062\pm 0.014            $ & $3.053\pm 0.014            $ \\
{\boldmath$n_\mathrm{s}   $} & $0.9702\pm 0.0032$ & $0.9670\pm 0.0035$  & $0.9734\pm 0.0038          $ & $0.9708\pm 0.0041          $ & $0.9797\pm 0.0041          $ & $0.9777\pm 0.0042          $ \\
{\boldmath$100\theta_\mathrm{MC}$} & $1.04106\pm 0.00023$ & $1.04084\pm 0.00023$ & $1.04227 \pm 0.00095$ & $1.04236\pm 0.00115$ & $1.0446\pm 0.00160$ & $1.0452 \pm 0.00175$ \\
{\boldmath$\Omega_\mathrm{b} h^2$} & $0.02229\pm 0.00011$ & $0.02227\pm 0.00011$ & $0.02245\pm 0.00015        $ & $0.02246\pm 0.00016        $ & $0.02271\pm 0.00014        $ & $0.02273\pm 0.00015        $ \\
{\boldmath$\Omega_\mathrm{c} h^2$} & $0.11820\pm 0.00080$ & $0.11827\pm 0.00080$ & $0.11850\pm 0.00082        $ & $0.11863\pm 0.00085        $ & $0.11831\pm 0.00094        $ & $0.11862\pm 0.00099        $ \\
{\boldmath$\tau_\mathrm{reio}$} & $0.0564\pm 0.0070$ & $0.0557\pm 0.0069$  & $0.0570\pm 0.0071          $ & $0.0560\pm 0.0071          $ & $0.0596\pm 0.0073          $ & $0.0588\pm 0.0073          $ \\
$H_0 {\rm [km/s/Mpc]}$ & $67.95\pm 0.35$ & $67.85\pm 0.35$ & $68.39^{+0.43}_{-0.50}$ & $68.38^{+0.47}_{-0.55}$ & $69.47\pm 0.51             $ & $69.56^{+0.50}_{-0.58}$ \\
$\Omega_\mathrm{m}$ & $0.3057\pm 0.0047$ & $0.3068\pm 0.0047$ & $0.3028\pm 0.0050          $ & $0.3032\pm 0.0052          $ & $0.2936\pm 0.0046          $ & $0.2935\pm 0.0046          $ \\
$\Omega_\mathrm{m} h^2$ & $0.14114\pm 0.00077$ & $0.14118\pm 0.00077$ & $0.14160\pm 0.00082        $ & $0.14174\pm 0.00086        $ & $0.14166\pm 0.00097        $ & $0.1420\pm 0.0010          $ \\
$\sigma_8$ & $0.8101\pm 0.0054$ & $0.8056\pm 0.0054$ & $0.8141\pm 0.0061          $  & $0.8100\pm 0.0064$ & $0.8194\pm 0.0070          $ & $0.8168\pm 0.0071          $ \\
$S_8$ & $0.8177\pm 0.0089$ & $0.8146\pm 0.0088$ & $0.8178\pm 0.0088          $ & $0.8142\pm 0.0089          $ & $0.8105\pm 0.0086          $ & $0.8079\pm 0.0084          $ \\
$r_\mathrm{drag} h \ {\rm [Mpc]}$ & $100.35\pm 0.62$ & $100.19\pm 0.61$ & $100.72\pm 0.67            $  & $100.65\pm 0.69            $ & $101.97\pm 0.64            $  & $101.97\pm 0.65            $ \\
$r_\mathrm{drag}  \ {\rm [Mpc]}$ & $147.67\pm 0.21$ & $147.67\pm 0.21$ & $147.27 \pm 0.34  $ & $147.19\pm 0.38 $ & $146.78\pm 0.46            $ & $146.59 \pm 0.50$ \\
$z_{\rm{drag}}$ & $1059.61\pm 0.25$ & $1059.58\pm 0.25$ & $1061.58 \pm 1.43$ & $1062.0\pm 1.68     $ & $1065.0\pm 2.23 $ & $1066.0\pm 2.44$ \\
$r_\mathrm{s} \ {\rm [Mpc]}$ & $144.97\pm 0.20$ & $144.97\pm 0.20$ & $144.60 \pm 0.32  $ & $144.53\pm 0.35$ & $144.17\pm 0.45 $ & $143.98 \pm 0.49$ \\
$z_\mathrm{s}$ & $1089.86\pm 0.18$ & $1089.89\pm 0.18$ & $1091.57 \pm 1.33 $ & $1092.02 \pm 1.60 $ & $1094.6\pm 2.33 $ & $1095.7 \pm 2.54$ \\
$M_b$ & - & - & - & - & $-19.378\pm 0.015$ & $-19.375^{+0.015}_{-0.017}$ \\
\hline
$\chi^2_{\rm highl}$ & $10546.6$ & $10550.6$ & $10546$ & $10545.6$ & $10549.4$ & $10553.4$ \\
$\chi^2_{\rm lensing}$ & $8.35$ & $8.89$ & $8.36$ & $8.78$ & $8.60$ & $8.68$ \\
$\chi^2_{\rm lowlTT}$ & $22.74$ & $23.03$ & $20.41$ & $20.90$ & $20.60$ & $20.19$ \\
$\chi^2_{\rm lowlEE}$ & $398.04$ & $396.023$ & $398.04$ & $395.78$ & $396.86$ & $398.99$ \\
$\chi^2_{\rm PL}$ & $10975.7$ & $10978.6$ & $10972.8$ & $10971.1$ & $10975.4$ & $10981.2$ \\
$\chi^2_{\rm DESI}$ & $15.15$ & $16.41$ & $14.36$ & $17.59$ & $12.74$ & $12.80$ \\
$\chi^2_{\rm ACT}$ & $242.16$ & - & $240.55$ & - & $243.12$ & - \\
$\chi^2_{\rm SPT}$ & - & $1877.4$ & - & $1878.5$ & - & $1882.2$ \\
$\chi^2_{\rm PL+DESI+ACT}$ & $11233.0$ & - & $11227.7$ & - & $11231.3$ & - \\
$\chi^2_{\rm PL+DESI+SPT}$ & - & $12872.4$ & - & $12867.2$ & - & 12876.2 \\
 \hline \\
\end{tabular}
\caption{\label{tab:params_actspt} 
The median values and the 68\% credible intervals of the {\bf primary} and selected  derived parameters of the $\Lambda$CDM and the $b\Lambda$CDM models fit to combinations of Planck (PL), DESI BAO, Pantheon+ (PP) SN, SH0ES-calibrated PP (PP+$M_b$), ACT DR4 and SPT-3G data, along with the best fit $\chi^2$ values.}
\end{table*}


%

\end{document}